\documentclass[12pt,english,american,aps, manuscript]{revtex4-1}
\usepackage{bera}

\usepackage[T1]{fontenc}
\usepackage[latin9]{inputenc}
\usepackage{geometry}
\usepackage[active]{srcltx}
\usepackage{array}
\usepackage{multirow}
\usepackage{amsmath}
\usepackage{graphicx}

\makeatletter

\providecommand{\tabularnewline}{\\}

\providecommand{\tabularnewline}{\\}

\usepackage{subfig}

\usepackage{babel}

\usepackage{footmisc}
\usepackage{hyperref}

\@ifundefined{showcaptionsetup}{}{%
 \PassOptionsToPackage{caption=false}{subfig}}
\usepackage{subfig}
\makeatother

\usepackage{babel}
\begin{document}

\title{A Java Application to Characterize Biomolecules and Nanomaterials
in Electrolyte Aqueous Solutions}

\author{Marcelo Marucho\footnote{Email: marcelo.marucho@utsa.edu, website: http://physics.utsa.edu/Faculty/Marucho/Marucho.html}}

\affiliation{Department of Physics and Astronomy, The University of Texas at San
Antonio, San Antonio, Texas 78249}
\begin{abstract}
The electrostatic, entropic and surface interactions between a macroion
(nanoparticle or biomolecule), surrounding ions and water molecules
play a fundamental role in the behavior and function of colloidal
systems. However, the molecular mechanisms governing these phenomena
are still poorly understood. One of the major limitations in procuring
this understanding is the lack of appropriate computational tools.
Additionally, only experts in the field with an extensive background
in programming, who are trained in statistical mechanics, and have
access to supercomputers are able to study these systems. To overcome
these limitations, in this article, \foreignlanguage{english}{we present
a free, multi-platform, portable Java software, which provides experts
and non-experts in the field an easy and efficient way to obtain an
accurate molecular characterization of electrical and structural properties
of aqueous electrolyte mixture solutions around both cylindrical-
and spherical-like macroions under multiple conditions. The Java software
does not require outstanding skills, and comes with detailed user-guide
documentation. The application is based on the so-called Classical
Density Functional Theory Solver (CSDFTS), which was successfully
applied to a variety of rod-like biopolymers, rigid-like globular
proteins, nanoparticles, and nano-rods. CSDFTS implements several
electrolyte and macroion models, uses different level of approximation
and takes advantage of high performance Fortran90 routines and optimized
libraries. These features enable the software to run on single processor
computers at low-to-moderate computational cost depending on the computer
performance, the grid resolution, and the characterization of the
macroion and the electrolyte solution, among other factors. As a unique
feature, the software comes with a graphical user interface (GUI)
that allows users to take advantage of the visually guided setup of
the required input data to properly characterize the system and configure
the solver. Several examples on nanomaterials and biomolecules are
provided to illustrate the use of the GUI and the solver performance. }
\end{abstract}
\maketitle
\selectlanguage{english}%

\section*{Program Summary:}

\noindent \emph{Program Title:} CSDFTS

\noindent \emph{Journal Reference:}

\noindent \emph{Catalogue identifier:}

\noindent \emph{Program obtainable from:} CPC Program Library; The
University at Texas at San Antonio, TX, USA. 

\noindent \emph{Licensing provisions:} UTSA license. 

\noindent \emph{Distribution format:} zip

\noindent \emph{Programming language:} Java 1.8.

\noindent \emph{Computer:} any computer at least with 2.8 GHz speed. 

\noindent \emph{Operating system:} Windows, Linux (Ubuntu, Fedora,
Centos, Debian), and Mac OSX. 

\noindent \emph{RAM:} It is required at least 6 GBs. We Recommend
12 GBs for highly demanding calculations.

\noindent \emph{Classification:} 3, 10. 

\noindent \emph{External routines/libraries:} GNU gawk installation
is required if the user wants to calculate the protein volume using
the 3v application. 

\noindent \emph{Subprograms used:} Jmol, pdb2pqr, Propka, Provol,
3v. 

\noindent \emph{Nature of problem:} A rich and complex, yet not fully
understood, electrical double layer (EDL) formation arises when a
nanoparticle or biomolecule is immersed in a liquid solution. 

\noindent \emph{Solution method:} The Java application is based on
the Classical Density Functional Theory Solver (CSDFTS).

\noindent \emph{Unusual features:} The software incorporates a graphical
user interface which eliminates the arduous and error-prone manual
entry of data, and substantially reduces the time spent on the setup
of the information required to characterize the macroion and the electrolyte
solution. Additionally, each GUI screen provides helpful information
about how to fill out the input data by simply holding the mouse pointer
over the corresponding text or blank box. the GUI tests all the input
data before running the CSDFTS to avoid the incorrect use of the software
and prevent meaningless results.

\noindent \emph{Additional comments:}\foreignlanguage{american}{ the
GUI requires a variety of the user's computer applications, which
are a part of the basic operating systems. The GUI uses ``Activity
Monitor'', ``procexp.exe'', and ``gnome-system-monitor'' (or
``top'') applications for Mac, Windows and Linux users, respectively,
to monitor the user's computer performance. It also uses ``Terminal.app'',
``cmd.exe'' and ``gnome-terminal'' (or ``xterm'') for Mac, Windows,
and Linux users, respectively, to display the CSDFT calculations.}

\noindent \emph{Running time:} the software to run on single processor
computers at low-to-moderate computational cost depending on the computer
performance, the grid resolution, and the characterization of the
macroion and the electrolyte solution.
\selectlanguage{american}%

\section{Introduction }

\selectlanguage{english}%
A rich and complex, yet not fully understood, electrical double layer
(EDL) formation arises when a nanoparticle or biomolecule is immersed
in a liquid solution \cite{key-1,key-9,key-2}. In such conditions,
the structural and thermodynamic properties of the liquid surrounding
the macroion are shown to be very different from those corresponding
to the bulk phase, which in turn, have a high impact on the behaviors
and functions of the macroion. For instance, physicochemical properties
of nanomaterials such as particle size (PS), shape, Zeta potential
(ZP), and surface charge density (SCD) have significant effects on
their stability, circulation in blood and absorption into cell membranes
\cite{key-3,key-4,key-5,key-6}. Alteration of the pH and electrolyte
conditions modify the SCD and ZP, which affect the binding of nanomaterials
to biological tissue and target them to specific sites inside the
cell. Whereas, in biophysics \cite{key-7,key-8,key-10}, the formation
of high order structures, including bundles and networks, involve
interactions between highly charged linker proteins and biopolymers
that are often modulated by the biological environment. The environment
has been shown to have a profound impact on biological functions.
For instance, ZP and SCD in microfilaments, nucleic acids, and globular
proteins play a fundamental role in the stabilization of these systems
due to charge residue groups on their surface. Extremely positive
or negative ZP values cause larger repulsive forces. This repulsion
between macroions with similar electric charge prevents aggregation
of the macroions and thus ensures easy redispersion. Adversely, altering
the ZP to the point at which it exhibits zero net charge may decrease
stabilization forces and promote agglomeration. 

Beyond the substantial progress done in the characterization of these
colloidal systems, the complex interplay between the physicochemical
properties of macroions and biological environment still remains elusive
due to the lack of appropriate methodologies. Conventional computational
tools and approaches are limited by their approximations and computational
cost. Certainly, the current understanding of EDL properties of macroions
is based mainly on mean-field theories like the non linear Poisson-Boltzmann
(NLPB) formalism and its modifications \cite{key-10}, which consider
electrostatic potential interactions only. Furthermore, the poor resolution
and sensitivity of current experimental techniques limit our ability
to obtain information on the molecular mechanisms governing these
phenomena. Thus, they may be inappropriate to explore a large number
of colloidal systems. Additionally, scientific software developed
to characterize solvation and electrical double layers of nanomaterials
and biomolecules might be helpful in understanding these phenomena.
However, they usually require specialized training and expertise in
computational biology, expensive commercial licenses, and access to
clusters or supercomputers, which are often an obstacle for many researchers,
experimentalists, even students lacking these requirements. Thus,
it is essential to develop not only more accurate and efficient approaches,
but also readily accessible, friendly implementation of a software
package that provides experts and non-experts a visualized guide (graphical
user interface) to perform these calculations without limitations
\cite{key-11}. 

\begin{figure}[h]
\selectlanguage{american}%
\begin{centering}
\includegraphics[scale=0.8]{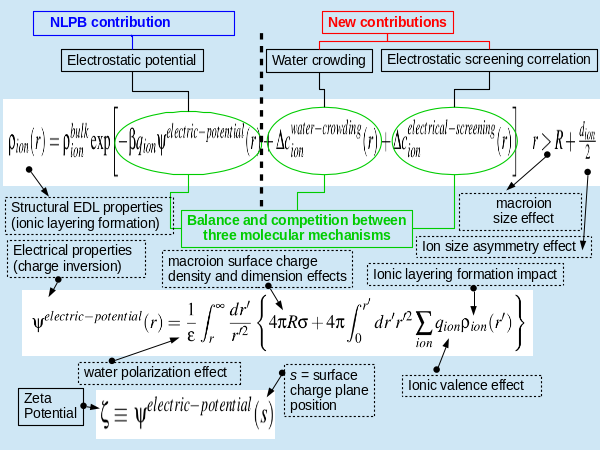}
\par\end{centering}
\selectlanguage{english}%
\caption{CSDFT theory}

\label{fig:CSDFT_theory}
\end{figure}

\selectlanguage{american}%
In this article, \foreignlanguage{english}{we present a free, multi-platform
and portable Java software which provides both experts and non-experts
in the field an easy and efficient way to get an accurate molecular
characterization of the EDL properties for biomolecules and nanomaterials
at infinite dilution. The application is based on the so-called Classical
Solvation Density Function Theory (CSDFT) and its modifications, which
have been shown to be particularly useful in studying multiple environmental
scenarios for a variety of rod-like\cite{key-12,key-13} and spherical\cite{key-14,key-15,key-16}
rigid-like macroions without computational restriction. CSDFT extends
the capabilities of NLPB formalism, eliminating the extremely high
computational demands of full atomistic simulation calculations without
losing important structural features of complex EDL properties (see
Figure \ref{fig:CSDFT_theory}). It considers not only the electric,
but also the entropic and many-body interactions. This feature has
been particularly useful for identifying and characterizing dominant
interactions and molecular mechanisms governing the behavior and function
of macroions under a variety of electrolyte conditions. }

\selectlanguage{english}%
In the next sections, we explain how to install, run and use the software.
We also include examples to illustrate the solver performance and
applicability. In Appendix A, we provide a brief summary of the approach
and computational scheme implemented by the software to solve the
CSDFT numerically. 
\selectlanguage{american}%

\section{Overview}

The software allows the user to use a variety of models and approaches.
It provides four macroion models: \emph{nanoparticle, nanorod, globular
protein,} and \emph{rod-like biopolymer}. For each of these models,
there are two approaches to characterize the macroion size and the
surface charge density: \emph{experimental} and \emph{protonation/deprotonation}
models for nanomaterials applications, and \emph{experimental} and
\emph{molecular structure} models for biophysics applications.\foreignlanguage{english}{
}Additionally, the software comes with two electrolyte approaches:
NLPB and CSDFT. These approaches include 3 ion size types (Crystal,
Effective and Hydrated), and\foreignlanguage{english}{ two solvent
models (implicit and explicit). As a unique feature, the Java application
comes with a graphical user interface (GUI) which eliminates the arduous
and error-prone manual entry of data, and substantially reduces the
time spent on the setup of the information required to characterize
the macroion and the electrolyte solution. Additionally, each GUI
screen provides helpful information about how to fill out the input
data by simply holding the mouse pointer over the corresponding text
or blank box. The GUI also provides default values for key input parameters
and preselects some relevant algorithms to speed up the setup of the
input data. However, they may be easily changed at any time. Moreover,
the GUI tests all the input data before running the CSDFTS to avoid
the incorrect use of the software and prevent meaningless results.
At the end of the calculations, the GUI generates two-dimension plots
of the selected output files to provide graphical representations
of the structural and electrostatic properties of the macroion EDL.
Finally, all the output data files are properly saved and organized
in the designated folder for post-analysis purposes. }

\section{General considerations}

\subsubsection{System Requirements}

The software requires at least 500 MB of disk space to install. Additional
disc space is required to save the output files generated by each
simulation, which may be as large as 1Gb or more. The available RAM
memory required to run the software depends on many factors. Based
on the examples presented in this article, most simulations require
only a few Gbs whereas highly demanding calculations may demand up
to 12 Gbs. 

The software is portable, e.g. it does not require external libraries,
leave its files or settings on the host computer, or modify the existing
system and its configuration. The software comes with Java 1.8 which
is configured by the installer to run the GUI. The Fortran90 codes
are distributed as binary files. For biophysics applications, the
software requires the molecular structure visualization java tool
``Jmol'' (http://jmol.sourceforge.net/) \cite{key-17} and either
the java based protein volume ``provol'' (gmlab.bio.rpi.edu/) \cite{key-26}
or the Unix based ``v3'' (http://3vee.molmovdb.org/volumeCalc.php)
\cite{key-18} applications. These open sources are compiled and distributed
along with the software. To use the latter application, the user's
computer needs to have the GNU ``gawk'' application installed (https://www.gnu.org/software
/gawk/). The software also includes ``pdb2pqr'' and ``propka''
binary files with the necessary libraries to perform molecular structure
calculations (http://apbs-pdb2pqr.readthedocs.io /en/latest/pdb2pqr/index.html)
\cite{key-19}. Additionally, the GUI requires a variety of the user's
computer applications, which are a part of the basic operating systems.
In particular, the GUI uses ``Activity Monitor'', ``procexp.exe'',
and ``gnome-system-monitor'' (or ``top'') applications for Mac,
Windows and Linux users, respectively, to monitor the user's computer
performance. It also uses ``Terminal.app'', ``cmd.exe'' and ``gnome-terminal''
(or ``xterm'') for Mac, Windows, and Linux users, respectively,
to display CSDFT calculations. 

\subsubsection{Installation}

\begin{figure}[h]
\begin{centering}
\includegraphics[scale=0.5]{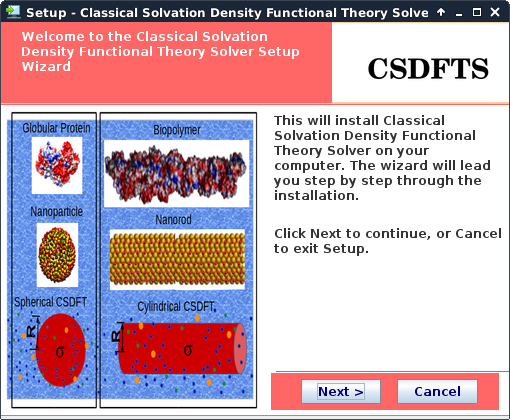}\includegraphics[scale=0.5]{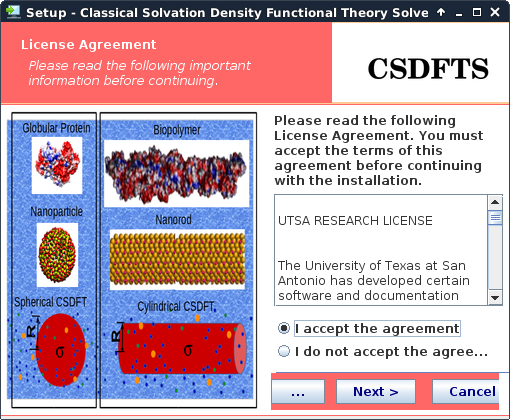}
\par\end{centering}
\begin{centering}
\includegraphics[scale=0.5]{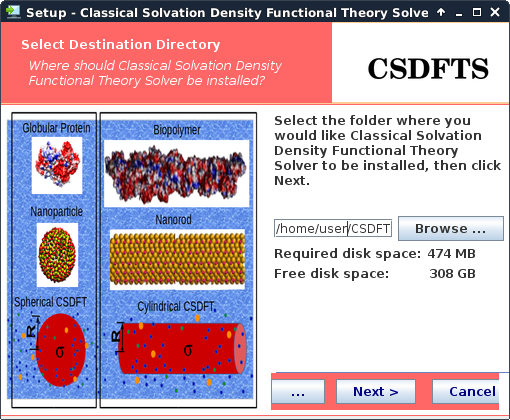}\includegraphics[scale=0.5]{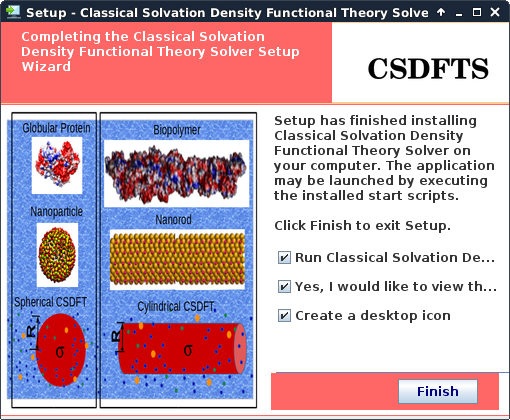}
\par\end{centering}
\caption{Multi-platform JAVA self-installer application}

\label{fig:installer}
\end{figure}

The software is distributed as a single zip file containing the self-installer
application. The user does not need ``Root'' or ``Admin'' permissions
to install the software on the user's home directory. The user may
unzip the file on the desktop directory, open the resulting folder,
then double-click the self-installer application to run the wizard
setup and follow a number of module steps (See Figure \ref{fig:installer}).
The installer creates the folder ``CSDFTS'' in the chosen directory
to save the software source files, including the GUI executable ``CSDFTS-software''.
If selected, the installer also creates an icon on the desktop for
easy access to run the GUI. The software was successfully installed
and tested on a variety of operating systems (Windows 7 and 10, Sierra,
Centos 7, Ubuntu 14, Fedora 24, Debian 8) and provides an uninstaller
application for easy removal.

\subsubsection{Output files}

\begin{figure}[h]
\begin{centering}
\includegraphics[scale=0.3]{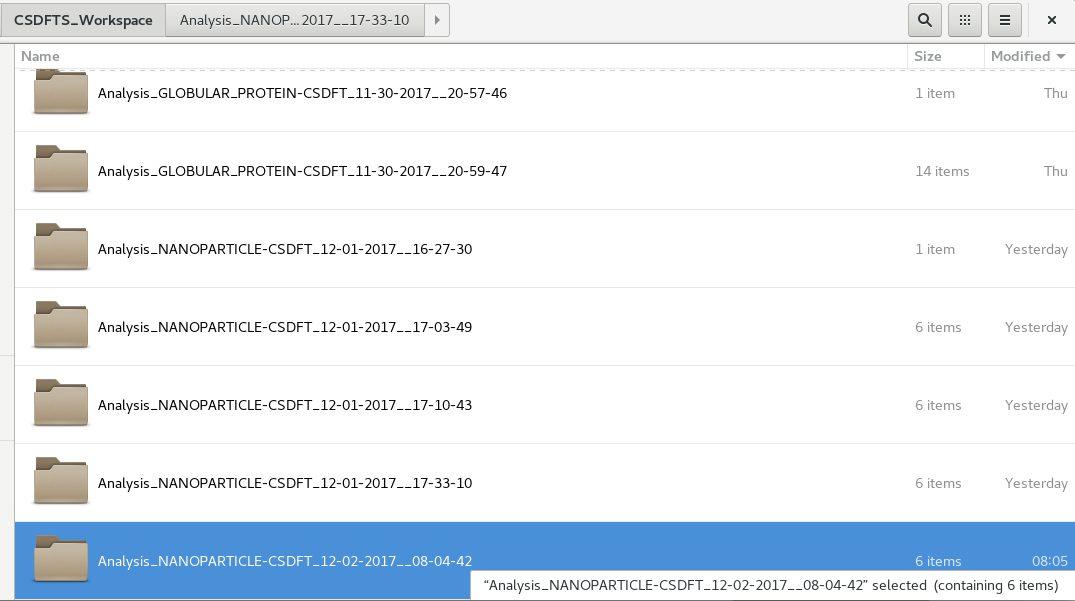}
\par\end{centering}
\begin{centering}
\includegraphics[scale=0.3]{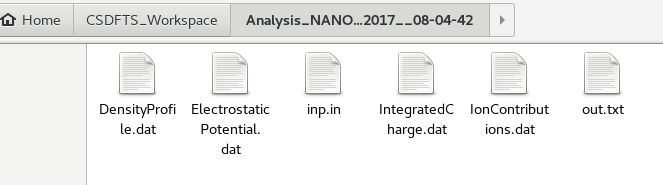}
\par\end{centering}
\begin{centering}
\includegraphics[scale=0.3]{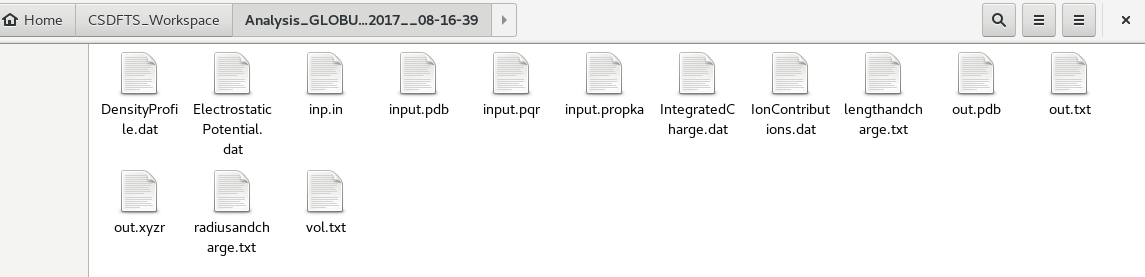}
\par\end{centering}
\caption{CSDFTS Workspace.}

\label{fig:CSDFTS_workspace}
\end{figure}

During the first simulation the GUI creates a folder named ``CSDFTS\_Workspace''
in ``My Documents'' and ``\$HOME'' directory for Windows and Unix
operating systems, respectively. Within this folder, the GUI creates
one sub-folder per each simulation to save and organize all the corresponding
output data files. The name of the sub-folders follows the convention: 

Analysis\_<\textcompwordmark{}<macroion-electrolyte>\textcompwordmark{}>\_<\textcompwordmark{}<Date>\textcompwordmark{}>\_\_<\textcompwordmark{}<Time>\textcompwordmark{}>,
where <\textcompwordmark{}<macroion- electrolyte>\textcompwordmark{}>
is the macroion and electrolyte model selected by the user, <\textcompwordmark{}<Date>\textcompwordmark{}>
is the current date, and <\textcompwordmark{}<Time>\textcompwordmark{}>
is the real time when the model was selected (see Figure \ref{fig:CSDFTS_workspace}). 

The GUI generates the following output files: ``DensityProfile.dat
(normalized ions and water density profile distributions''); ``ElectrostaticPotential.dat''
(mean electrostatic potential); ``IntegrateCharge.dat'' (the integrated
charge), and ``IonContributions.dat''(normalized electrostatic potential
energy, particle crowding entropy energy, and ion-ion electrostatic
correlation energy to the ionic potential of mean force). Explicit
expressions for these properties are provided in Appendix A. The size
of these files may vary depending on the number of ions species, number
of grid points, and electrolyte model. These files are formatted in
a multi-column arrangement where the row number corresponds to the
number of grid points, the first column contains the discretized distance
in unit of $\textrm{Å}$ and the remaining column(s) provide the numerical
solution(s) (see Table \ref{tbl:output_files}). \foreignlanguage{english}{Other
relevant data including SCD, ZP, and PS are saved in the log file
``out.txt'', whereas} information on the input data and solver configuration
are included in the files ``inp.in'' and ``input.file'' for spherical
and cylindrical macroions, respectively. If the \textit{molecular
structure} option is selected for biophysical applications, the GUI
generates the following additional files: ``input.pdb'' (copy of
the molecular structure uploaded by the user); ``input.pqr'' (pdb2pqr
output data); ``input.propka'' (propka output data) ; ``out.pdb''
(molecular structure information with heterogem atoms removed); ``out.xyzr''
(atomic positions in xyzr format); ``vol.txt'' (3v and provol output
data including information on the macroion surface and volume as well
as the number of residues and atoms); ``lengthandcharge.txt'' (total
macroion charge and dimensions), and ``radiusandcharge.txt'' (parameters
characterizing the macroion radius and charge density). 

\begin{table}[h]
\begin{tabular}{|c|c|c|}
\hline 
File name & Units & Description\tabularnewline
\hline 
\hline 
DensityProfile.dat &  & $\begin{array}{c}
\textrm{The first column is for distances. The second column corresponds to}\\
\textrm{the first ion species selected in ion model section, and so on. }\\
\textrm{If the CSDFT is selected, the last column corresponds to water}
\end{array}$\tabularnewline
\hline 
ElectrostaticPotential.dat & KT/e & $\begin{array}{c}
\textrm{The first column is for distances.}\\
\textrm{the second column contains the values of the mean }
\end{array}$\tabularnewline
\hline 
IntegratedCharge.dat & e & $\begin{array}{c}
\textrm{The first column is for distances. }\\
\textrm{the second column contains the values of the integrated charge}
\end{array}$\tabularnewline
\hline 
IonConstributions.dat & J/KT & $\begin{array}{c}
\textrm{The first column is for distances. Then, there is one set of columns per}\\
\textrm{contribution. The number of columns per contribution is equal to the}\\
\textrm{number of ion species. The order of the set of columns }\\
\textrm{from left to right reads : the electrostatic potential }\\
\textrm{energy, excluded volume energy, the ion-ion correlation energy, }\\
\textrm{and the potential of mean force (sum of the previous contributions). }\\
\textrm{The order of the columns in each set of contributions is the following:}\\
\textrm{first column corresponds to the first selected ion species, and so on.}
\end{array}$\tabularnewline
\hline 
\end{tabular}

\caption{Output Files created by CSDFTS.}

\label{tbl:output_files}
\end{table}

\section{Graphical User Interface: Use and Applications}

\subsection{GUI Description}

\begin{figure}[h]
\begin{centering}
\includegraphics[scale=0.6]{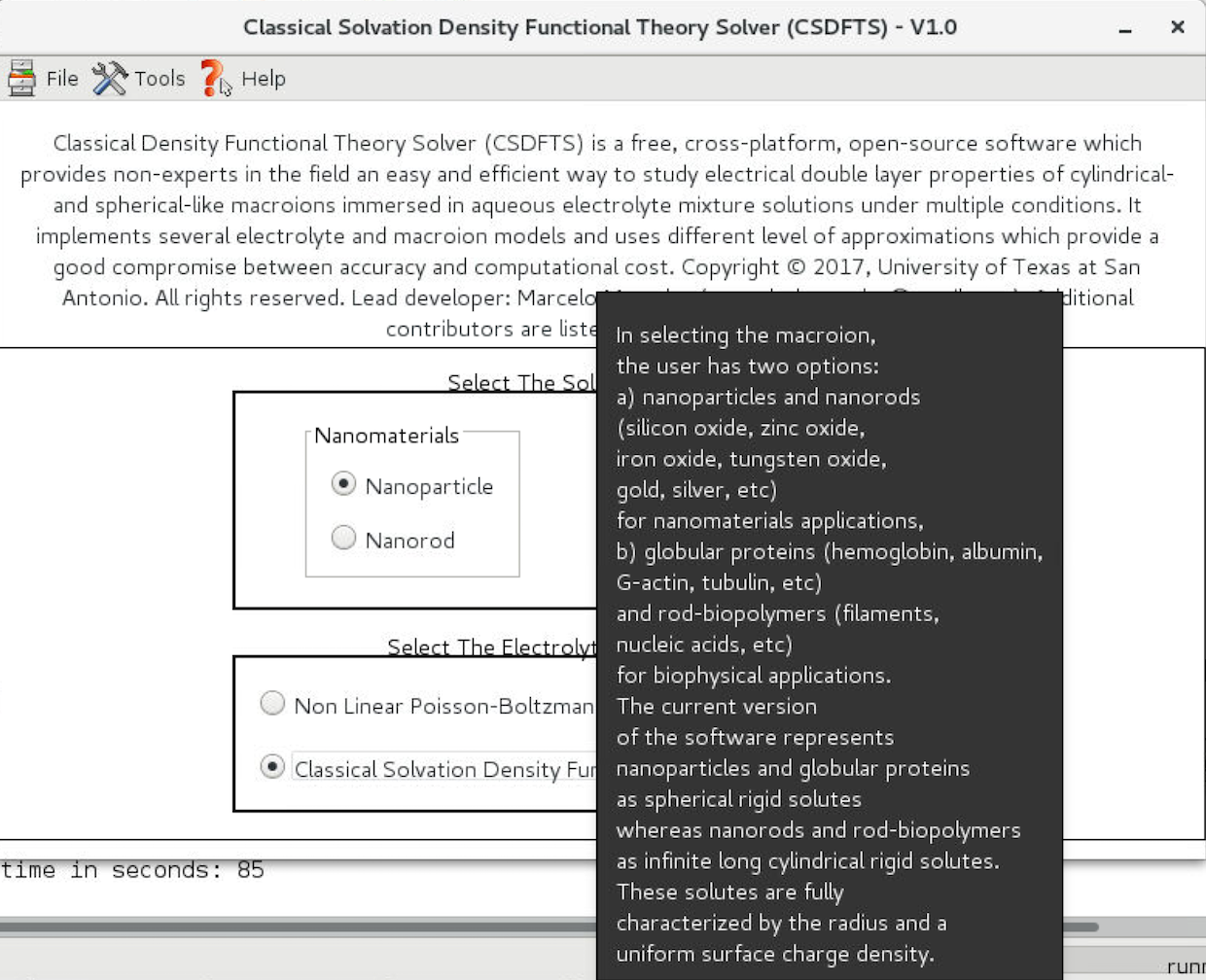}
\par\end{centering}
\caption{Help Messages.}

\label{fig:help-message}
\end{figure}

In this section we describe the screen sequence generated by the GUI.
The first, second and third screens correspond to ``\emph{project
window}'', ''\emph{model window}'', and ``\emph{results visualization
window}'', respectively. \foreignlanguage{english}{Each screen provides
information to help the user fill out the input data by moving the
mouse pointer over the corresponding text or blank box (see Figure
\ref{fig:help-message}).}

\subsubsection{Project Window: Selection of the solute geometry model and electrolyte
solution theory}

\begin{figure}[h]
\begin{centering}
\includegraphics[scale=0.5]{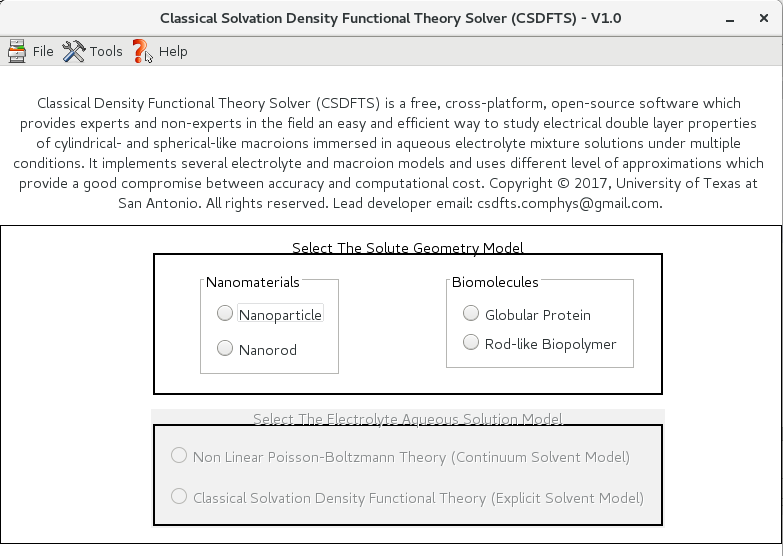}
\par\end{centering}
\caption{Project Window - main screen. }

\label{fig:mainscreen}
\end{figure}

The Project window shown in Figure \ref{fig:mainscreen} is the main
screen which provides user access to the \textit{Menu}, \textit{Information}
and \textit{Model Sections}. 

\begin{figure}[h]
\begin{centering}
\includegraphics[scale=0.6]{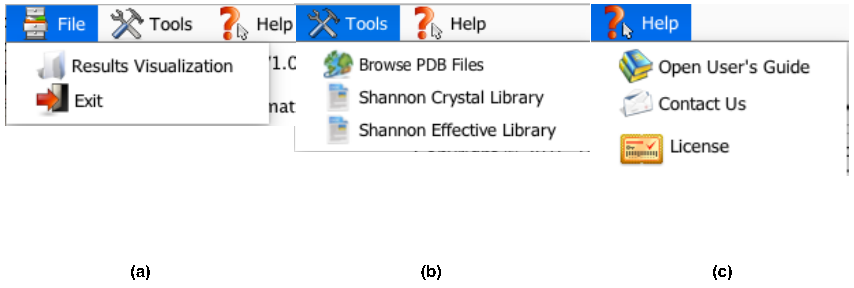}
\par\end{centering}
\caption{(a) File Menu. (b) Tools Menu. (c) Help Menu.}

\label{fig:menus-1-1}
\end{figure}

\textit{The menu section}, located at the top left corner of the window,
contains the \textit{File}, \textit{Tools} and \textit{Help menu}s.
The \textit{File Menu} contains the \textit{Results Visualization}
and \textit{Exit} options (see Figure \ref{fig:menus-1-1}(a)). The
\textit{Results Visualization} option may be used after running the
simulation. This option allows users to select output file(s) and
visualize the solution(s) in two dimensional plot(s). 

The \textit{Tools Menu} contains the \textit{Browse PDB files}, \textit{Crystal
Ion Library, Effective Ion Library} and Hydrated Ion Library options
(see Figure \ref{fig:menus-1-1}(b)). The tool \textit{Browse PDB
files} is used for biophysical applications only. It allows users
to open a web browser and download protein molecular structure(s)
from the protein data base web page (http://www.rcsb.org/). The user
may use this tool at anytime. The \textit{Ion Library} tools provide
the user access to the tables pretabulated with specific information
on ion species, valences and diameters that are required to characterize
the electrolyte aqueous solution. These tools can be used to redefine
existing ion species and define new ion species as well. The last
menu in Figure \ref{fig:menus-1-1}(c) is the \textit{Help Menu} which
provides access to the user's guide, software license and contact
information.

The \textit{Solute Geometry Model} section is located at the central-bottom
part of the window. To select the macroion (solute), the user has
four options: \textit{Nanoparticles} and \textit{nanorods} for nanomaterials
applications, or \textit{globular proteins} and \textit{rod-biopolymers}
for biophysical applications. 

The \textit{Electrolyte Aqueous Solution Model} is located right below
the \textit{Solute geometry model}. CSDFTS offers two electrolyte
aqueous solution theories. NLPB uses an implicit solvent model and
considers electric interactions only, whereas, CSDFT uses an explicit
solvent model (SPM) and considers not only the electric, but also
the entropic and ion-ion correlation interactions. More information
on solvent and ion models is provided in Appendix A. The selection
of the EDL model mainly depends on the computational resources, the
electrolyte conditions, physicochemical properties of the solute,
and the accuracy required for the numerical solution (illustrative
examples are provided below). As a rule of thumb, NLPB is more efficient,
but less accurate than CSDFT. It is worth mentioning that CSDFT is
a unique feature of the software, whereas NLPB theory is included
for testing and comparison purposes, mostly. Indeed, there are other
efficient programs based on implicit solvent models such as APBS \cite{key-20}
and MPBEC \cite{key-21} who provide the solution for both symmetric
and asymmetric macroion shapes.

\subsubsection{Model Window: system and solver configuration}

\begin{figure}[h]
\begin{centering}
\subfloat[Nanomaterials screen]{\begin{centering}
\includegraphics[scale=0.35]{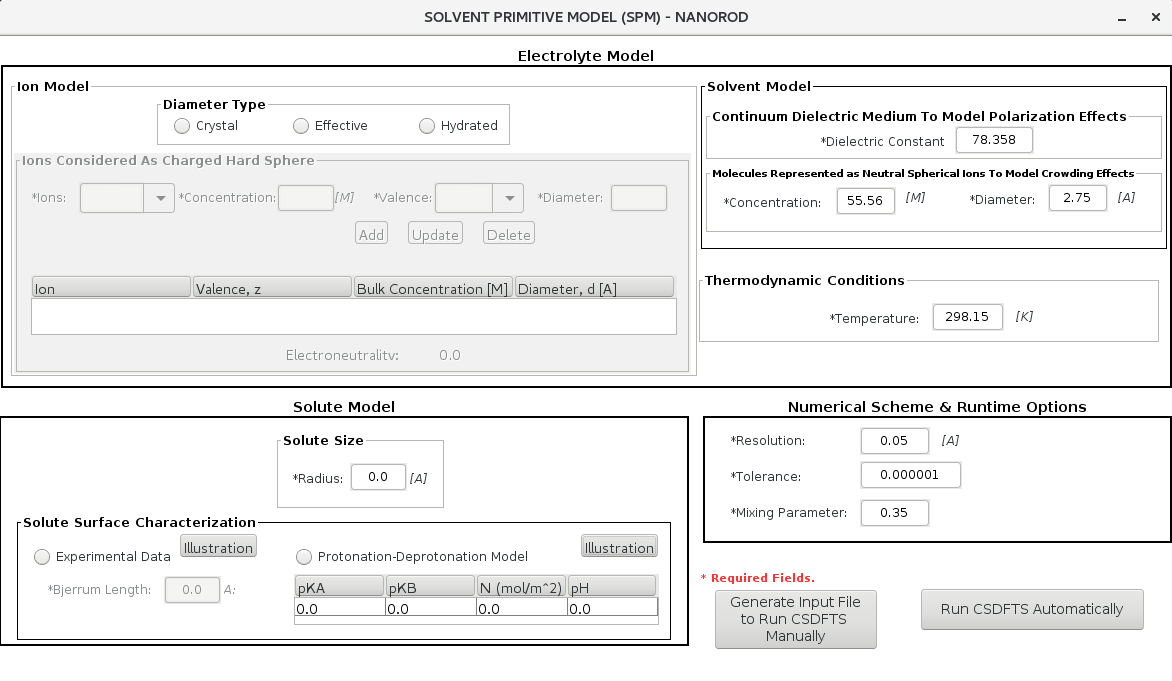}
\par\end{centering}
}
\par\end{centering}
\begin{centering}
\subfloat[Biophysics screen]{\begin{centering}
\includegraphics[scale=0.35]{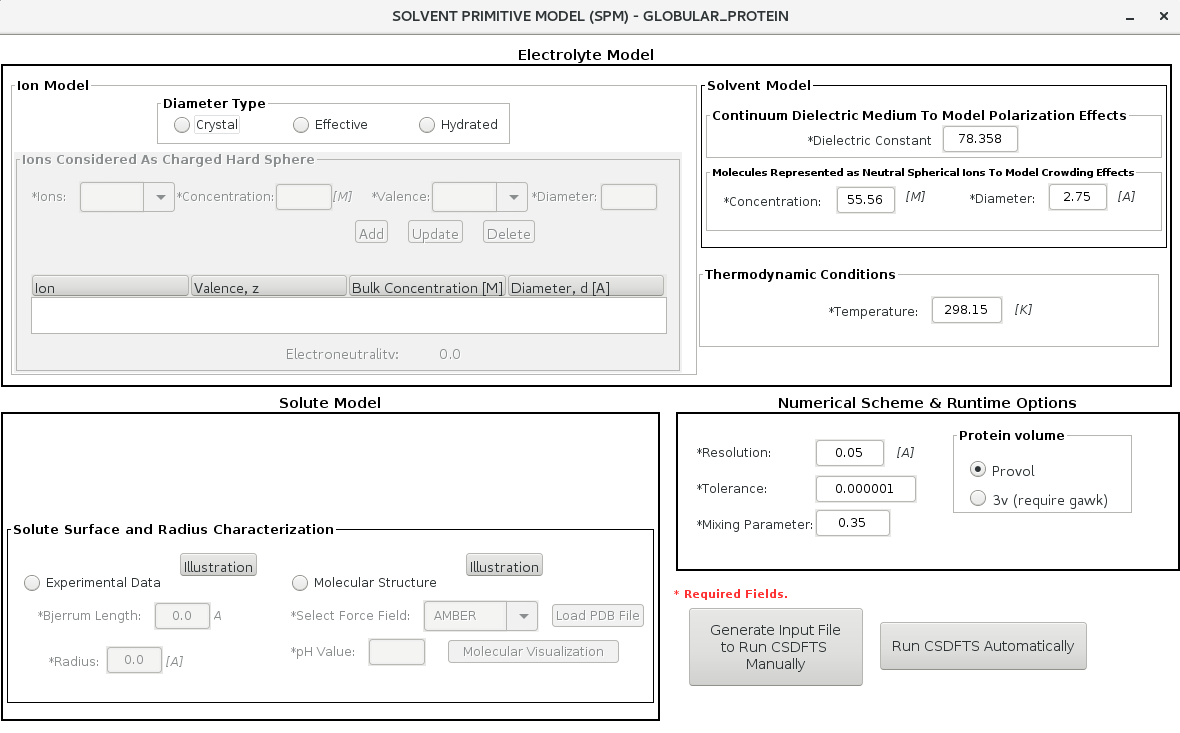}
\par\end{centering}
}
\par\end{centering}
\caption{Model window option}

\label{fig:NLPB}
\end{figure}

Once the \textit{Solute Geometry Model} and the \textit{Electrolyte
Aqueous Solution Model} are selected in the \textit{project window},
the GUI automatically opens a second screen, the \textit{model window}.
In this screen, most of the text based user interaction occurs (see
Figure \ref{fig:NLPB}). It contains the following modules:\textit{
Electrolyte} Model, \textit{Solute Model}, and\textit{ Numerical Scheme
and Runtime }\textit{\emph{Options}}. 

\begin{figure}[h]
\begin{centering}
\includegraphics[scale=0.5]{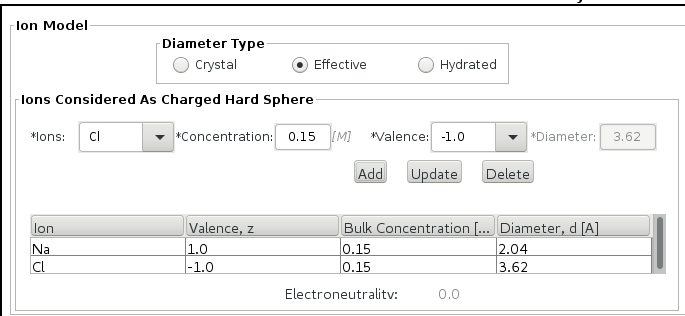}
\par\end{centering}
\caption{Ion Model.}

\label{fig:IonModel}
\end{figure}

The \textit{Ion Model} section shown in Figure \ref{fig:IonModel}
allows users to characterize the electrolyte solution by providing
\foreignlanguage{english}{the bulk concentration $[\rho_{i}^{0}]$
and the valence $z_{i}$} for each ion species $i$. In NLPB theory,
\foreignlanguage{english}{ions are represented as point-like particles,
and consequently, there is no need to define the ion sizes. }Whereas,
in CSDFT \foreignlanguage{english}{each ionic species $i$ is represented
by charged hard spheres of diameter $d_{i}$.} In this case, CSDFTS
offers a pretabulated crystal, effective and hydrated ionic diameter
types, which are estimated using different experimental techniques
\cite{key-22}. To characterize the electrolyte solution, the user
has to select the first ion species and its properties, then click
the button ``Add'', and subsequently repeat the procedure to add
more ion species. The ``Delete'' and ``Update'' buttons allow
users to remove and change the properties of a previously selected
ion. Once the user selects all the ion species (e.g. valence and bulk
concentration), the electroneutrality condition (e.g. $\sum_{i}z_{i}[\rho_{i}^{0}]=0$)
must be satisfied. The status of this condition is displayed at the
bottom of this section. Note that only ion species comprising the
electrolyte solution must be defined in this section. Hydrogen and
Hydroxide ions controlling the pH level of the electrolyte solution
are assigned by CSDFTS automatically. 

\begin{figure}[h]
\begin{centering}
\includegraphics[scale=0.5]{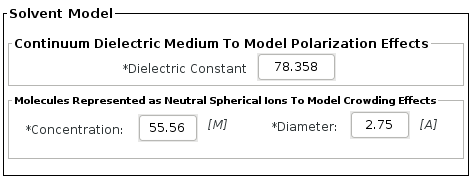}
\par\end{centering}
\caption{Solvent Model.}

\label{fig:solventmodel}
\end{figure}

The \textit{Solvent Model section} shown in Figure \ref{fig:solventmodel}
allows users to characterize the structural and electric properties
of the solvent. Both NLPB and CSDFT consider the solvent as neutral
polar molecules. NLPB only requires the value of the uniform bulk
dielectric permittivity constant \foreignlanguage{english}{$\epsilon$}
to model the solvent electrical properties. CSDFT additionally requires
the solvent molar bulk concentration \foreignlanguage{english}{$[\rho_{w}^{0}]$}
and molecular diameter $d_{w}$ to model the solvent entropy properties.\foreignlanguage{english}{
}The default values displayed in this section correspond to the experimental
values for water. More details on solvent models are provided in Appendix
A.

\begin{figure}[h]
\begin{centering}
\subfloat[Nanomaterials]{\begin{centering}
\includegraphics[scale=0.5]{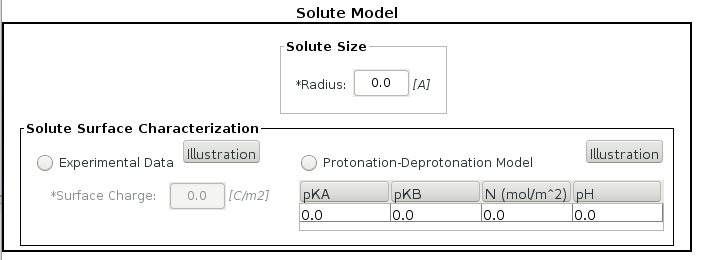}
\par\end{centering}
}
\par\end{centering}
\begin{centering}
\subfloat[Biomolecules]{\begin{centering}
\includegraphics[scale=0.5]{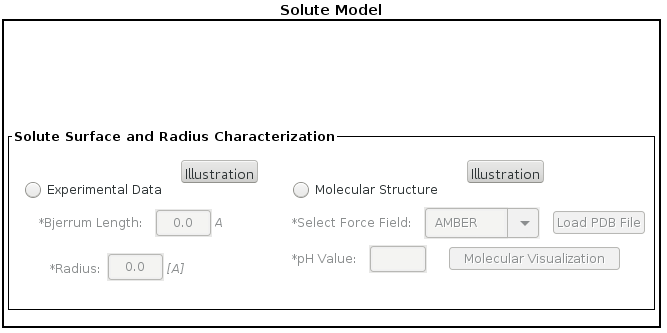}
\par\end{centering}
}
\par\end{centering}
\caption{Solute Model.}

\label{fig:solutemodel}
\end{figure}

Regarding the \textit{Solute Model section,} Figure \ref{fig:solutemodel}
a) corresponds to nanomaterials, whereas Figure \ref{fig:solutemodel}
b) corresponds to biomolecules. The current version of the software
represents nanoparticles and globular proteins as spherical rigid
solutes. Nanorods and rod-biopolymers are represented as infinitely
long cylindrical rigid solutes. These solutes are characterized by
the effective radius, as well as, the surface charge density $\sigma$
and the Bjerrum length $\ell$ for spherical and cylindrical shapes,
respectively.

For nanomaterials it is required to provide the effective radius which
is usually obtained from experiments. For biomolecules there are two
ways to characterize the solute size. The user may either provide
this information from experiments (\textit{Experimental data model})
or it can be estimated from the molecular structure (\textit{Molecular
structure model}).

\begin{figure}[h]
\begin{centering}
\includegraphics[scale=0.4]{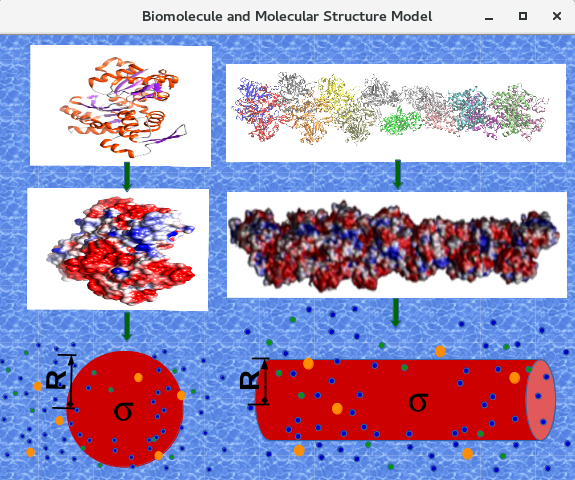}
\par\end{centering}
\begin{centering}
\includegraphics[scale=0.4]{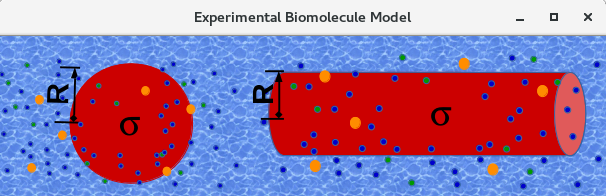}
\par\end{centering}
\begin{centering}
\includegraphics[scale=0.4]{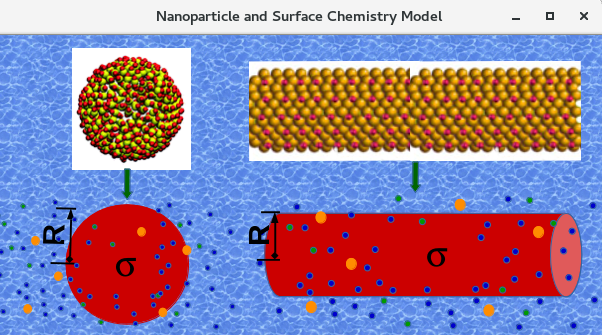}
\par\end{centering}
\caption{Illustrations}

\label{fig:illustrations}
\end{figure}

\begin{figure}[h]
\begin{centering}
\includegraphics[scale=0.5]{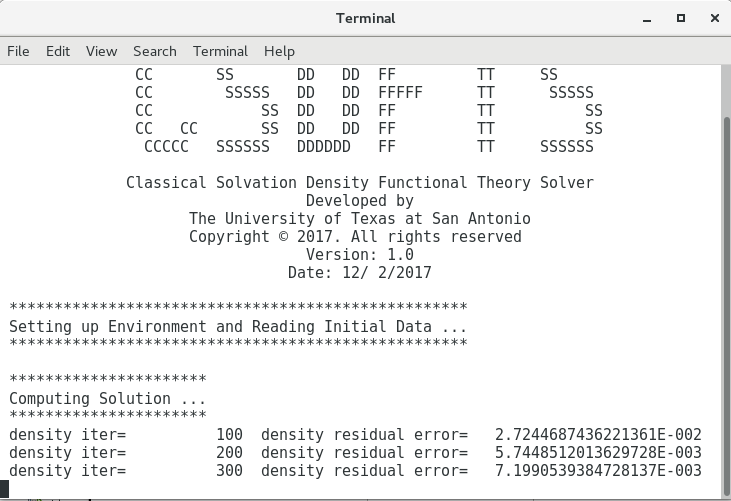}
\par\end{centering}
\caption{CSDFTS calculations}

\label{fig:terminal}
\end{figure}

\begin{figure}[h]
\begin{centering}
\includegraphics[scale=0.5]{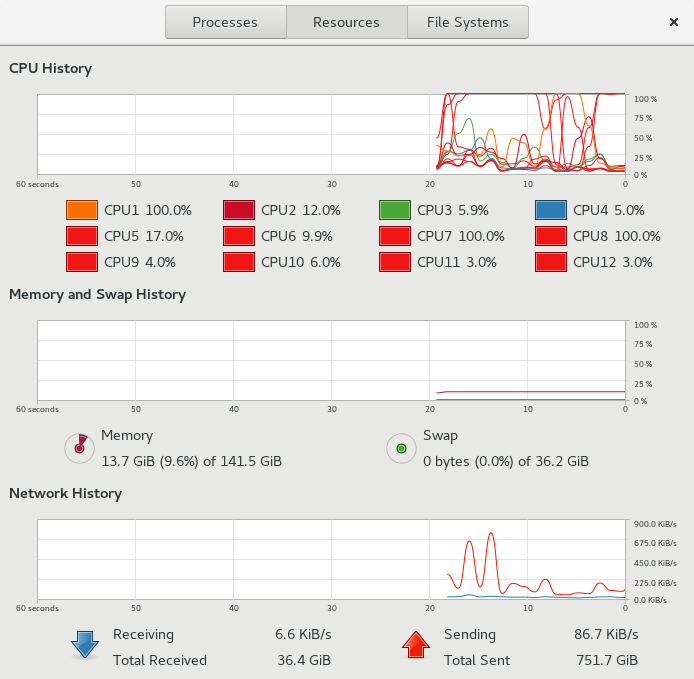}
\par\end{centering}
\caption{User's computer performance on Linux operating systems}

\label{fig:RAMmemory}
\end{figure}

A novel feature of the software is the option to calculate the macroion
surface charge density $\sigma$ arising from the protonation/deprotonation
reactions of the dissociable functional groups at the solid/liquid
interface. The user can also click the \emph{Illustration} buttons
to visualize the different options and models (see Figure \ref{fig:illustrations}).
For nanomaterials, this charging / discharging mechanism has traditionally
been described using surface complexation models (SCMs), where titration
of surface groups is described by the pH level and an ensemble of
mass balance equilibria with associated equilibrium constants (see
Appendix A). This approach is included in the \textit{protonation/deprotonation
model} where the user has to provide the equilibrium constants \textit{pKa}
and \textit{pKb} for protonation and deprotonation of the active functional
group, respectively, as well as the number density of total functional
groups on the surface \textit{N}, and the \textit{pH} level (see Figures
\ref{fig:solutemodel} a)). The resulting SCD depends on the ZP and
vise-versa. A different scenario is presented for globular proteins
and rod-biopolymers since the titration is usually described by a
set of empirical rules relating the protein structure to the equilibrium
constant values of ionizable groups (residues). This mechanism is
included in the\textit{ Molecular structure model} option where the
user has to upload the molecular structure (pdb file), select a force
field, and define the pH level as shown in Figure \ref{fig:solutemodel}
b). This uncharged molecular structure in pdb format is used by the
software to automatically run the pdb2pqr / propka application \foreignlanguage{english}{(http://sourceforge.net/projects/pdb2pqr/).
This application} assigns atomic charges and sizes, adds missing hydrogens,
optimizes the hydrogen bonding network, and renormalizes atomic charges
of the residues exposed to the surface due to pH effects (protonation/deprotonation
process). Afterwards, the GUI uses the molecular structure in pdb
format to automatically run\foreignlanguage{english}{ either the provol
(default) or 3v application to }estimate the total macroion volume
by rolling a probe particle of radius 1.4$\textrm{Å}$ and using a
grid mesh of 0.5$\textrm{Å}$. Finally, the GUI uses the information
obtained from the total protein charge and volume to estimate the
effective solute radius as well as the uniform surface charge density
and Bjerrum length for spherical and cylindrical shapes, respectively.
For each of these approaches, the GUI opens an external terminal window
to display these calculations \foreignlanguage{english}{(see Figure
\ref{fig:terminal})}. It also opens a RAM memory window to monitor
the user's computer performance (see Figure \ref{fig:RAMmemory}).
The GUI closes these windows once the calculations are over and enables
the execution buttons \textit{``Generate the input file and run CSDFT
manually''} and \textit{``run CSDFT automatically''} to run CSDFT. 

\begin{figure}[h]
\begin{centering}
\includegraphics[scale=0.5]{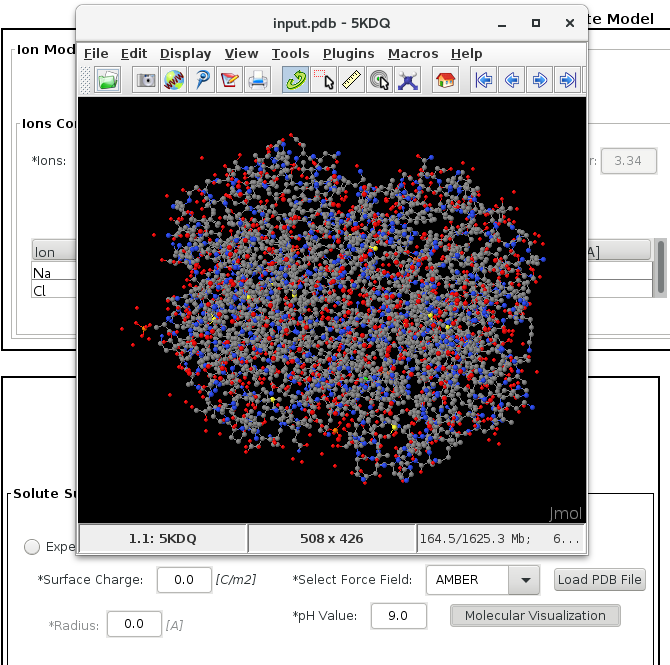}
\par\end{centering}
\caption{Molecular structure visualization.}

\label{fig:jmol-1}
\end{figure}

\selectlanguage{english}%
Another important feature for biophysics applications is the \emph{molecular
visualization} tool. After all the aforementioned calculations are
over, the user can visualize the uploaded molecular structure which\foreignlanguage{american}{
provides a detailed molecular characterization including the amino
acid sequence and the number and type of residues exposed to the electrolyte
}(see Figure \ref{fig:jmol-1})\foreignlanguage{american}{. Alternatively,
the user may select the \textit{experimental data} option to provide
the values of these parameters from experiments. In this case, there
is no waiting time to run CSDFTS. }

\selectlanguage{american}%

\begin{figure}[h]
\begin{centering}
\includegraphics[scale=0.4]{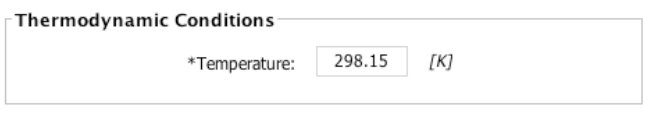}
\par\end{centering}
\caption{Electrolyte solution temperature.}

\label{fig:thermocond}
\end{figure}

The \textit{Thermodynamic Conditions} option is used to define the
electrolyte solution temperature. The default value displayed in Figure
\ref{fig:thermocond} corresponds to the room temperature. 

\begin{figure}[h]
\begin{centering}
\includegraphics[scale=0.6]{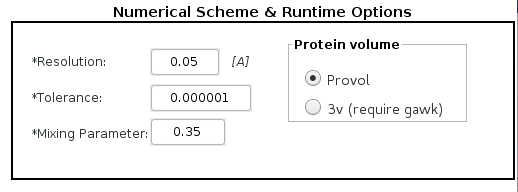}
\par\end{centering}
\caption{Numerical scheme and runtime options.}

\label{fig:numericalscheme}
\end{figure}

The\textit{ Numerical Scheme and Runtime Option} section shown in
Figure \ref{fig:numericalscheme} allows the user to configure key
parameters that play a fundamental role in the solver performance.
They control the accuracy and computational cost. The \textit{tolerance}
number represents the numerical error required by the user to numerically
obtain the normalized density profile solutions. A minimum of six
digits of precision is highly recommended (default value). The radial
grid \textit{resolution} represents the regular separation distance
\textit{h} between two consecutive points in the domain discretized
of the radial distance to solve CSDFT numerically. The value recommended
for this parameter is 0.05$\textrm{Å}$ which has been shown to work
for most applications on computers without RAM memory restrictions.
The domain of the solution ranges from the macroion surface (e.g.
the radius \textit{R)} to the cutoff \textit{L.} The latter is determined
automatically by CSDFTS to provide the correct (long range) asymptotic
behavior of the mean electrostatic potential, and consequently, satisfy
the electroneutrality condition of the system. The number of grid
points is calculated as follows $N=(L-R)/h+1$. The \textit{Mixing
Parameter} is a number between 0 and 1 which helps the solver to reach
stability and convergence in the solution iteratively. The default
value for this parameter is 0.35. It is required for the \textit{protonation/deprotonation
model} only. 

We note that the high \emph{resolution} and \emph{tolerance generate}
high computational cost, allocated RAM memory, solver stability and
accuracy, but slow the rate of convergence on the calculations. The
user may use lower \emph{resolution} to speed up the calculations
and reduce the allocated RAM memory at the risk of loosing convergence
in the CSDFTS iterations. In that case, the solver will stop the simulation
after 10000 iterations. The user may also kill the process using standard
approaches (e.g. ``Ctrl+c'', etc). More details on the numerical
solver scheme are provided in Appendix A. 

\begin{figure}[h]
\begin{centering}
\includegraphics[scale=0.5]{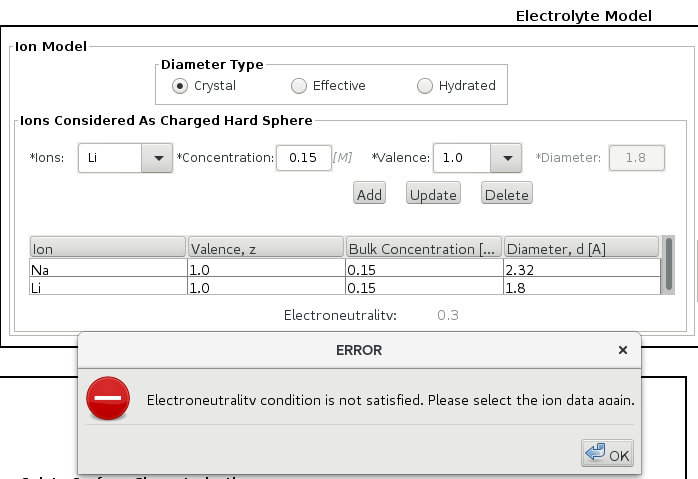}
\par\end{centering}
\caption{Warning messages.}

\label{fig:warning-message}
\end{figure}

After all changes on preselected parameters and options are completed,
all boxes are filled out and precalculations are over, the user may
either press the button \textit{Generate the input file and run CSDFT
manually} or \textit{run CSDFT automatically}. T\foreignlanguage{english}{he
GUI tests all the input data, and sends the user warning messages
if unusual/nonphysical values are assigned to the required parameters
or missing information is detected (see Figure \ref{fig:warning-message}).}

\selectlanguage{english}%
We highly recommend the user to use the option \foreignlanguage{american}{\textit{run
CSDFT automatically}}. The other option is mainly for those users
that have computer skills and prefer to run the simulations using
command line. In this case, the GUI generates the\foreignlanguage{american}{
input file ``inp.in'' for spherical solutes and ``inputfile.in''
for cylindrical solutes. To run CSDFTS manually in Linux operating
systems, for instance, the user should open a terminal, change the
current directory to the corresponding output sub-folder directory
and type ``\textasciitilde{}/CSDFT/Linux/sphere\_linux.sh inp.in''
for spherical solutes or ``\textasciitilde{}/CSDFT/Linux/cylinder\_linux.sh
inputfile.in'' for cylindrical solutes. A similar procedure should
be performed for other operating systems.}

\selectlanguage{american}%

\subsubsection{Results Visualization}

\begin{figure}[h]
\begin{centering}
a)\includegraphics[scale=0.3]{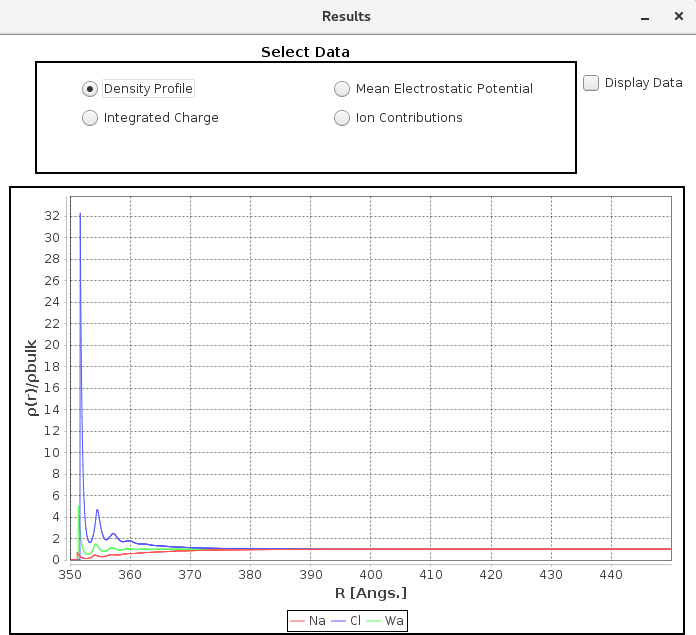}b)\includegraphics[scale=0.3]{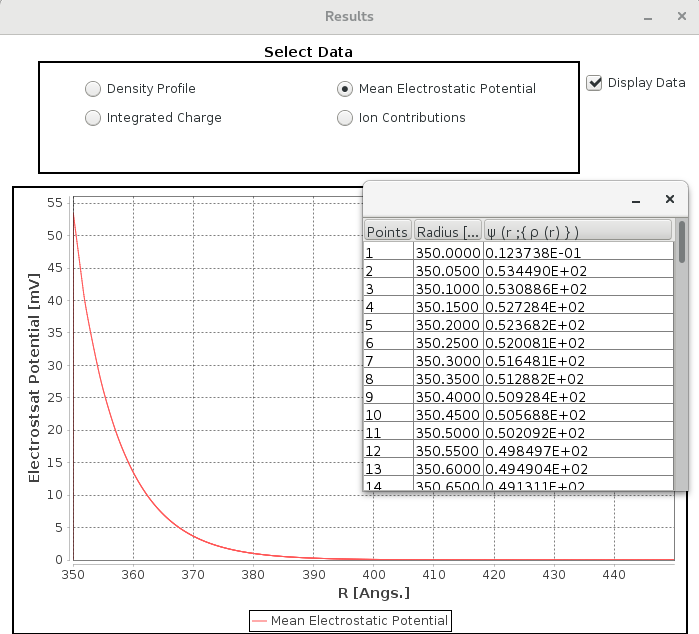}
\par\end{centering}
\begin{centering}
c)\includegraphics[scale=0.3]{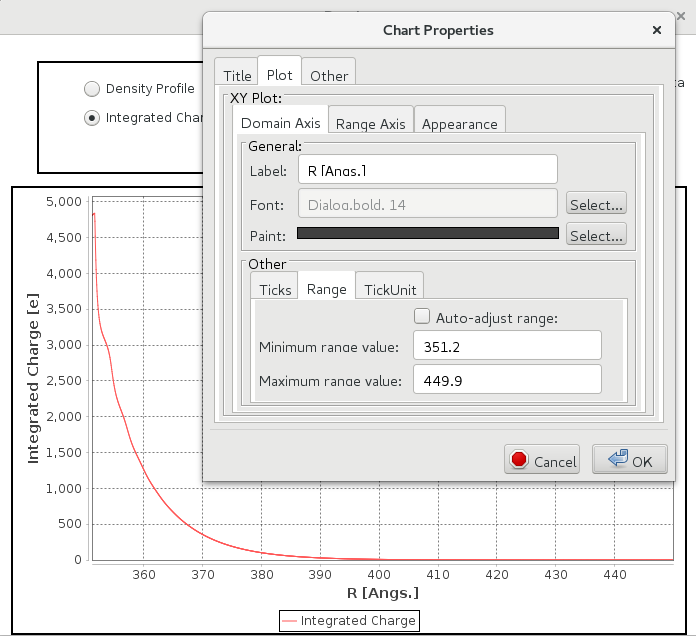}d)\includegraphics[scale=0.3]{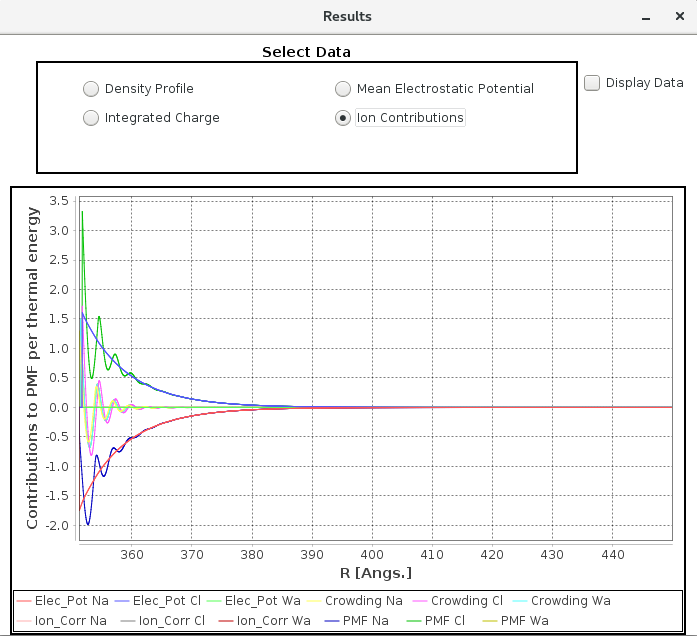}
\par\end{centering}
\caption{Results Visualization}

\label{fig:openoutput-1-1}
\end{figure}

If the \textit{``run CSDFT automatically''} option is selected,
the \textit{results visualization} window appears once the simulation
is over. This screen allows the user to select the output file(s)
and visualize the numerical solution(s) as a function of the radial
separation distance in two dimension plot(s) (see Figures \ref{fig:openoutput-1-1}).
To change the plot properties the user must right click on the desired
plot (see Figure \ref{fig:openoutput-1-1} c)). The user can also
visualize the data on spreadsheets as shown in Figure \ref{fig:openoutput-1-1}
b). 

\subsection{Quick start}

The user is recommended to follows these simple steps:
\begin{itemize}
\item On the first screen, select the macroion type and subsequently the
electrolyte model. 
\item On the second screen, select the ions and provide the corresponding
concentrations. Make sure to fulfill the electroneutrality condition.
Next, define the solvent model. The dielectric permittivity value
may be changed if needed. If CSDFT was selected, the solvent concentration
and molecule size may be also changed. Subsequently, configure the
solver. The resolution and tolerance may be changed depending on the
user's computer performance and available RAM memory. The default
values may be used if there is no RAM memory restrictions. Otherwise,
decrease the resolution accordingly. For nanomaterials applications
and protonation / deprotonation models, the mixing parameter may be
changed if needed. For biophysics applications and the molecular structure
option, the protein volume application may be changed depending on
the user's operating system. The default application works for all
the operating systems whereas ``3v'' only works for Unix operating
systems and requires the GNU gawk application installed on the user's
computer. As a rule of thumb, the latter runs faster and requires
less allocated RAM memory. Next, the temperature may be changed if
needed. The last step is the selection of the solute model. For nanomaterials
applications, select either the experimental or protonation / deprotonation
option and provide the corresponding information. Now the software
is ready to run the CSDFTS. For biophysics applications, select either
the experimental or the molecular structure option and provide the
corresponding information. In the latter case, select the force field
and set up the pH before uploading the molecular structure. Wait until
some calculations on the molecular structure are over. Afterwards,
the molecular structure my be visualized. Finally, select either the
\textit{``Generate the input file and run CSDFT manually'' }\textit{\emph{or
}}\textit{``run CSDFT automatically''.}
\item If the \textit{``run CSDFT automatically''} was selected, an additional
screen is available to visualize solutions in 2D plots.
\item If there is any issue in the previous steps, close all screens and
start over again. Double check that all the input data and selected
options are appropriate for the required simulation and change those
parameters to improve the CSDFTS convergence and the user's computer
performance. 
\end{itemize}

\subsection{Examples\label{sec:Examples}}

In this section we analyze several macroions under different electrolyte
conditions (see Table \ref{table:examples2}). We ran these simulations
on several operating systems and CPU processors. The results shown
in the present article corresponds to those obtained on a single Intel
Xeon 5680 and CentOS 7 operating system.

The first example is a Spherical Silica oxide nanoparticle of radius
50$\textrm{Å}$ immersed in a salt mixture of 0.2M \emph{$NaCl$}
+ 0.05 M $MgCl_{2}$ (e.g. 0.2M $Na^{+}$, 0.05 M \emph{$Mg^{+2}$},
and 0.3M \emph{$Cl^{-}$} after complete dissociation), with protonotation/deprotonation
defined by equilibrium constants \emph{pKa} = 6.8 and \emph{pKb} =1.7,
density number of active site \emph{N} = 0.000002$mol/m^{2}$, and
\emph{pH} 4 . The second example is a Silica oxide nanorod of radius
30$\textrm{Å}$ immersed in a single salt of 0.65 M \emph{$KCl$}
(e.g. 0.65 M \emph{$K^{+}$} and 0.65 M \emph{$Cl^{-}$}), with the
same values for \emph{pKa, pKb}, and \emph{N}, but \emph{pH} is set
equal to 11. The third example is a segment of B-DNA with experimental
value for the radius of 9.5$\textrm{Å}$ and Bjerrum length of 7.2$\textrm{Å}$
immersed in a mixed salt of 0.3M \emph{$NaCl$} +0.15M \emph{$KCl$}
(0.3 M $Na^{+}$, 0.15 M $K^{+}$, and 0.45 M $Cl^ {}$) at neutral
\emph{pH}. The fourth example is the myoglobin globular protein immersed
in a single salt of 0.15 M $CaCl_{2}$ (e.g. 0.15 M $Ca^{+2}$ and
0.3 M $Cl^{-}$), using the molecular structure 4of9.pdb (153 residues),
Amber force field and \emph{pH} 8. The last example is a filament
actin immersed in 0.35M \emph{$KCl$} (e.g. 0.35M $K^{+}$ and 0.35
M $Cl^{-}$), using the molecular structure 3B5U.pdb (375 residues),
Amber force field, and \emph{pH} 5. We use predefined values for the
remaining parameters unless otherwise is defined in table \ref{table:examples2}.
We also use the crystal ion type for the CSDFT approach. 

\begin{table}[h]
\begin{centering}
\begin{tabular}{|c|c|c|c|c|c|c|c|c|c|c|c|}
\hline 
\multirow{3}{*}{$h$$[\textrm{Å}]$} & \multirow{3}{*}{Properties} & \multicolumn{2}{c|}{Example} & \multicolumn{2}{c|}{Example} & \multicolumn{2}{c|}{Example} & \multicolumn{2}{c|}{Example} & \multicolumn{2}{c|}{Example}\tabularnewline
\cline{3-12} 
 &  & \multicolumn{2}{c|}{1} & \multicolumn{2}{c|}{2} & \multicolumn{2}{c|}{3} & \multicolumn{2}{c|}{4} & \multicolumn{2}{c|}{5}\tabularnewline
\cline{3-12} 
 &  & CSDFT & NLPB & CSDFT & NLPB & CSDFT & NLPB & CSDFT & NLPB & CSDFT & NLPB\tabularnewline
\hline 
\hline 
\multirow{2}{*}{} & pH & \multicolumn{2}{c|}{4} & \multicolumn{2}{c|}{11} & \multicolumn{2}{c|}{7} & \multicolumn{2}{c|}{8} & \multicolumn{2}{c|}{5}\tabularnewline
\cline{2-12} 
 & Radius $[\textrm{Å}]$ & \multicolumn{2}{c|}{50} & \multicolumn{2}{c|}{30} & \multicolumn{2}{c|}{9.5} & \multicolumn{2}{c|}{16.38} & \multicolumn{2}{c|}{22.13}\tabularnewline
\hline 
\multirow{4}{*}{$0.05$} & $\sigma\:[C/m^{2}]$ or $\ell\:[\textrm{Å}]$ & -0.0135 & -0.0137 & -0.1913 & -0.191 & 7.2 & 7.2 & 0.0048 & 0.0048 & 18.39 & 18.39\tabularnewline
\cline{2-12} 
 & ZP $[mV]$ & -8.59 & -12.14 & -3.94 & -93.66 & 7.81 & 19.48 & 1.44 & 2.65 & 1.69 & 4.13\tabularnewline
\cline{2-12} 
 & $\begin{array}{c}
\textrm{Max.RAM memory}\\{}
[Gb]
\end{array}$ & 2.9 & 1 & 2.5 & 1.5 & 1.1 & 0.5 & 0.7 & 0.7 & 11.5 & 11.5\tabularnewline
\cline{2-12} 
 & $\begin{array}{c}
\textrm{Approx. computing}\\
\textrm{time \ensuremath{[min]}}
\end{array}$ & 15 & <1 & 27 & <1 & 4 & <1 & 2 & 1 & 49 & 47\tabularnewline
\hline 
\multirow{4}{*}{$0.1$} & $\sigma\:[C/m^{2}]$ or $\ell\:[\textrm{Å}]$ & -0.0137 & -0.0137 & -0.193 & -0.191 & 7.2 & 7.2 & 0.0048 & 0.0048 & 18.39 & 18.39\tabularnewline
\cline{2-12} 
 & ZP $[mV]$ & -8.65 & -12.13 & -3.89 & -93.45 & 7.77 & 19.31 & 1.28 & 2.63 & 1.57 & 4.11\tabularnewline
\cline{2-12} 
 & $\begin{array}{c}
\textrm{Max.RAM memory}\\{}
[Gb]
\end{array}$ & 2.3 & 0.7 & 2.2 & 1.5 & 0.4 & 0.4 & 0.5 & 0.5 & 11.5 & 11.5\tabularnewline
\cline{2-12} 
 & $\begin{array}{c}
\textrm{Approx. computing}\\
\textrm{time \ensuremath{[min]}}
\end{array}$ & 9 & <1 & 19 & <1 & 1 & <1 & 2 & 1 & 51 & 47\tabularnewline
\hline 
 &  &  &  &  &  &  &  &  &  &  & \tabularnewline
\hline 
\end{tabular}
\par\end{centering}
\caption{Illustrative examples}

\label{table:examples2}
\end{table}

\begin{figure}[h]
\begin{centering}
a)\includegraphics[scale=0.25]{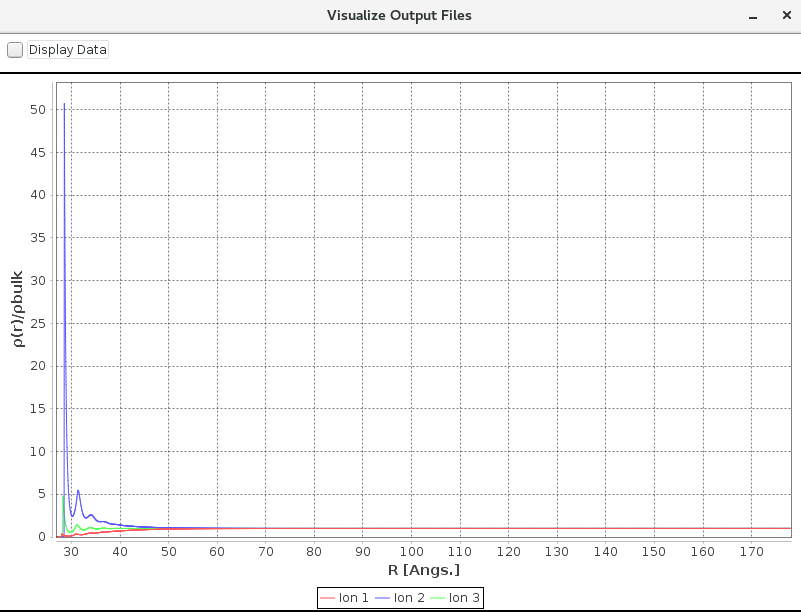}b)\includegraphics[scale=0.25]{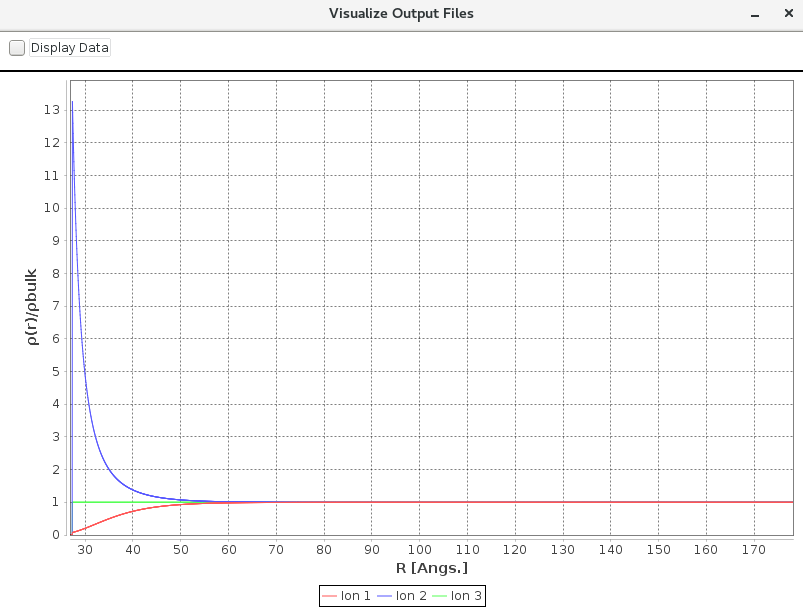}
\par\end{centering}
\caption{Example4: Ionic and water density distribution. Left and right plots
correspond to CSDFT and NLPB, respectively.}

\label{fig:ex4}
\end{figure}

\begin{figure}[h]
\begin{centering}
a)\includegraphics[scale=0.25]{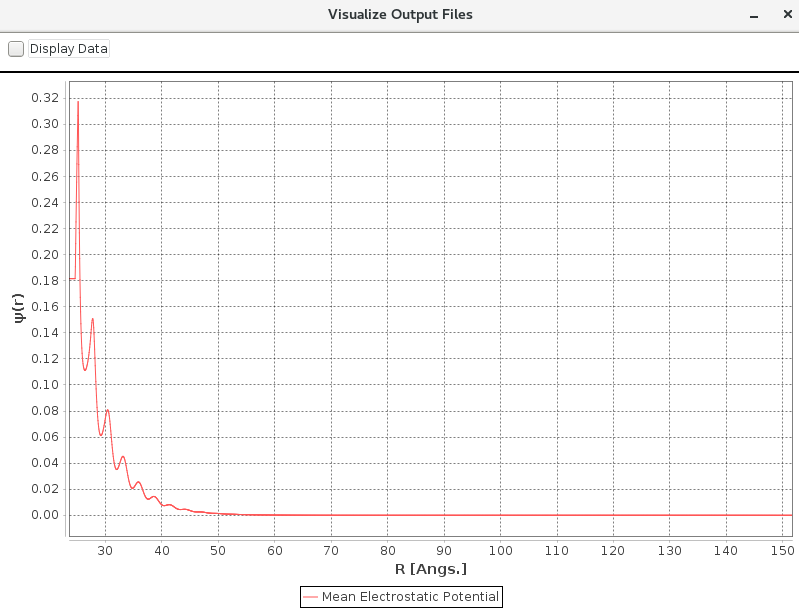}b)\includegraphics[scale=0.25]{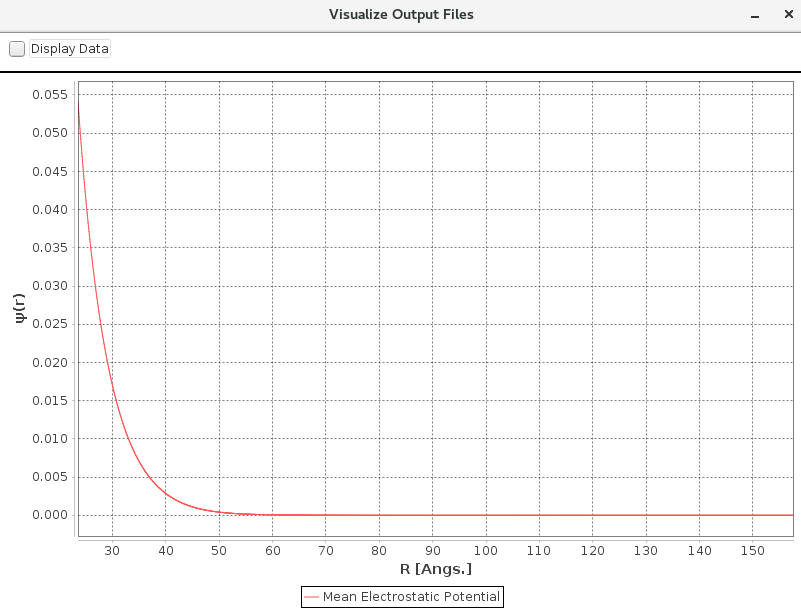}
\par\end{centering}
\caption{Example5: Mean electrostatic potential. Left and right plots correspond
to CSDFT and NLPB, respectively.}

\label{fig:ex5}
\end{figure}

To illustrate numerical solutions, we show in Figure \ref{fig:ex4}
the structural properties of the EDL predicted by CSDFT and NLPB for
a myoglogin protein, whereas in Figure \ref{fig:ex5}, we present
the electric properties of the EDL corresponding to a actin filament
. \foreignlanguage{english}{Certainly, this visualization is useful
for analyzing the behavior of the electrostatic potential interactions
and the ion-water distributions at short and long distances from the
macroion. It is also relevant for analyzing the effective range, intensity,
and nature of the driving force governing the potential of mean force
for each ion species.} An extensive analysis and discussion on structural
and electrical properties of EDLs can be found elsewhere \cite{key-10,key-12,key-14,key-15,key-16,key-23}. 

Overall, the software offers diversity in applicability and analysis
on many colloidal systems. For instance, running multiple simulations
at different pH levels may provide a molecular understanding on the
impact of pH on the ZP, SCD, and EDL properties of macroions, and
consequently, their stability and aggregation. Furthermore, simulating
different electrolyte solutions may be useful to understand the effects
of biological fluids on biomolecules and their functions. It is also
possible to perform multiple simulations for different macroion sizes,
charges and shapes to study their influence on membrane absorption. 

\section{Future Directions}

Users will be notified to update the software when a new release is
available. Additional features to be included in coming releases are:
1) modeling more complex charging mechanism on the surface of nanomaterials
including more than one active functional group and the release /adsorption
of ions; 2) modeling solvent polarization effects \cite{key-23}.
Future work will be focused on modeling: 1) inhomogenous surface charge
density and solutes with irregular shapes; 2) nanomaterials having
surface functionalization; 3) ionic currents along biopolymers\cite{key-13}.
\begin{acknowledgments}
This work was supported by NIH Grant 1SC2GM112578-03. The author warmly
thanks technicians Esteban Valderrama and Samrita Neogi for their
participation in programming and testing the software.
\end{acknowledgments}

\appendix

\section*{APPENDIX A: Theory and Numerical Scheme}

\subsection{CSDFT for Spherical and Cylindrical macroions immersed in a aqueous
electrolyte solution at neutral pH.}

\selectlanguage{english}%
In this approach, we consider a rigid charged macroion of effective
radius $R$ and uniform surface charge density $\sigma$ surrounded
by an electrolyte solution comprised of $m$ ionic species. We use
the solvent particle model to characterize the electrolyte. Each ionic
species $i$ is represented by bulk Molar concentration $[\rho_{i}^{0}]$,
a charged hard sphere of diameter $d_{i}$, and total charge $q_{i}=ez_{i}$,
where $e$ is the electron charge and $z_{i}$ is the corresponding
ionic valence. Additionally, the solvent molecules are represented
as a neutral ion species whereas the solvent electrostatics is considered
implicitly by using the continuum dielectric environment with a dielectric
constant $\epsilon$. The macroion-liquid interaction induces inhomogeneous
ion profiles $\left[\rho_{i}(r)\right]$ which are calculated using
CSDFT as follows \cite{key-12,key-14}:

\begin{align}
\left[\rho_{i}(r)\right]=\Bigg\{ & \begin{aligned}\left[\rho_{i}^{0}\right]exp\{\Delta E_{i}(r,\{[\rho_{j}])\}, & r>R+d_{i}/2\\
0, & r\leq R+d_{i}/2
\end{aligned}
\label{eq:ionprofilefinal}
\end{align}
where $\Delta E_{i}(r,\{[\rho_{j}])\equiv-\beta q_{i}\psi(r,\{[\rho_{j}]\})+\Delta c_{i}^{(1)hs}(r;\{[\rho_{j}]\})+\Delta c_{i}^{(1)res}(r;\{[\rho_{j}]\})$
stands for the ionic PMF per unit of thermal energy $KT$, $\beta=1/kT$
, $k$ is the Boltzmann constant, $T$ the temperature, and $c_{i}^{(1)hs}(r;\{[\rho_{j}]\})$
and $c_{i}^{(1)res}(r;\{[\rho_{j}]\})$ are the hard sphere (particle
crowding) and residual electrostatic ion-ion correlation functions,
respectively. $\psi(r,\{[\rho_{j}]\})$ represents the MEP of the
system

\begin{equation}
\psi(r,\{[\rho_{j}]\})=\frac{e}{\mbox{\ensuremath{\epsilon}}}\int_{r}^{\infty}\frac{dr'}{r'}P(r',\{[\rho_{j}]\},n)\label{eq:psi_def}
\end{equation}

and

\begin{equation}
P(r',n)=\frac{1}{r'^{n-1}}\left\{ \frac{R^{n}\sigma}{e}+\int_{R}^{r'}dr'r'^{n}\sum_{i}z_{i}\rho_{i}(r')\right\} \quad\left\{ \begin{array}{c}
\textrm{n=1 for cylindrical macroions}\\
\textrm{n=2 for spherical macroions}
\end{array}\right.\label{eq:intcharge}
\end{equation}

the integrated charge (see Figure \ref{fig:CSDFT_theory}). Expression
(\ref{eq:psi_def}) is the formal solution of the PB equation for
an homogeneous anisotropic dielectric media $\epsilon$

\begin{equation}
\begin{array}{c}
\nabla^{2}\psi(r,\{[\rho_{j}]\})=-\frac{1}{\epsilon}\sum_{i=1}^{m}z_{i}\left[\rho_{i}(r)\right]\}\\
\\
\epsilon\partial\psi(r,\{[\rho_{j}]\})/\partial r|_{r=s}=-\sigma,\qquad\psi(r,\{[\rho_{j}]\})|_{r\rightarrow\infty}\rightarrow0,
\end{array}\label{eq:poissoneq}
\end{equation}
 with the surface charge layer position defined as $s\equiv R+<\{d_{i}\}>$
and $<\{d_{i}\}>\equiv N_{A}l_{B}^{3}\sum_{i}\left|z_{i}\right|[\rho_{i}^{0}]d_{i}/(2m)$.
In the latter definition $N_{A}$ and $l_{B}$ stand for the Avogadro
number and the Bjerrum length, respectively.

\subsection{Surface Complexation Model: \foreignlanguage{american}{Accounting
for pH Effects of the Electrolyte aqueous Solution on the macroion
Surface Charge Density}}

In order to account for the titration that regulates the macroion
surface charge density $\sigma$ we consider the following two protonation
reactions of single \textit{MO}-coordinated sites :

\begin{equation}
MOH\leftrightarrow MO^{-}+H^{+},\qquad MOH+H^{+}\leftrightarrow MOH_{2}^{+}.\label{eq:chemicalreactions}
\end{equation}
with equilibrium constants $K_{A}$ and $K_{B}$ 

\begin{equation}
K_{A}=\frac{N_{MO^{-}}[H^{+}]_{s}}{N_{MOH}},\qquad K_{B}=\frac{N_{MOH_{2}^{+}}}{N_{MOH}[H^{+}]_{s}}.\label{eq:equilibriumconstants}
\end{equation}

In the above expressions, $N_{MOH}$, $N_{MO^{-}}$, and $N_{MOH_{2}^{+}}$
are the surface site densities of $MOH$, $SO^{-}$ and $MOH_{2}^{+}$,
respectively. $[H^{+}]_{s}$ is the concentration of $H^{+}$ ions
at the SCD position $s$, namely 

\begin{equation}
[H^{+}]_{s}\equiv\left[\rho_{H}(s)\right]=[\rho_{H}^{0}]exp\{\Delta E_{H}(s,\{[\rho_{j}])\},\label{eq:protonation}
\end{equation}
where $\Delta E_{H}(s,\{[\rho_{j}])=-\beta\zeta+\Delta c_{H}^{(1)hs}(s;\{[\rho_{j}]\})+\Delta c_{H}^{(1)res}(s;\{[\rho_{j}]\})$
is the hydrogen PMF per unit of thermal energy $KT$, $\zeta\equiv\psi(s,\{[\rho_{j}]\})$
is the ZP, and $\Delta c_{H}^{(1)hs}$ and $\Delta c_{H}^{(1)res}$
represent the hydrogen hard sphere (particle crowding) and ion-ion
correlation contributions, respectively. The bulk concentration of
$H^{+}$ ions is represented by $[\rho_{H}^{0}]$, which is related
to the $pH$ value of the bulk liquid at infinite dilution by the
expression $pH=-Log([\rho_{H}^{0}])$. The total number density of
functional groups on the SCD position is $N_{total}=N_{MO^{-}}+N_{MOH}+N_{MOH_{2}^{+}}$
and the SCD is $\sigma=-F(N_{MO}-N_{MOH_{2}^{+}})$, where $F$ represents
the Faraday constant . Writing the densities sites in terms of the
equilibrium constants (\ref{eq:protonation}), the SCD can be calculated
as follows \cite{key-15,key-16}:

\begin{equation}
\sigma=-FN_{total}\frac{K_{A}-K_{B}\left[\left[\rho_{H}(r)\right]_{r=s}\right]^{2}}{K_{A}+\left[\rho_{H}(r)\right]_{r=s}+K_{B}\left[\left[\rho_{H}(r)\right]_{r=s}\right]^{2}}\label{eq:complexation}
\end{equation}

The $pH$ of the solution is adjusted by adding strong acid ($NaOH$)
and ($HCl$) solutions to the electrolyte. The free proton and hydroxyl
ion bulk concentrations are given by the well-known expressions $[\rho_{H}^{0}]=10^{-pH},\quad and\quad[\rho_{OH}^{0}]=10^{-(14-pH)},$
respectively, and the bulk concentrations of the electrolyte are chosen
to satisfy the bulk electroneutrality condition.

\subsection{NLPB approach}

NLPB is a particular case of the CSDFT approach. Indeed, the expressions
introduced in previous sections for CSDFT recover the NLPB approach
by setting all ion sizes equal to zero. In particular, $s=R$, $\Delta c_{i}^{(1)res}(r;\{[\rho_{j}]\})=0,$
and $\Delta c_{i}^{(1)hs}(r;\{[\rho_{j}]\})=0$ in continuum models.
\selectlanguage{american}%

\subsection{Iterative Solver Scheme}

\begin{figure}[h]
\begin{centering}
\label{fig: solverscheme}\foreignlanguage{american}{\includegraphics[scale=0.6]{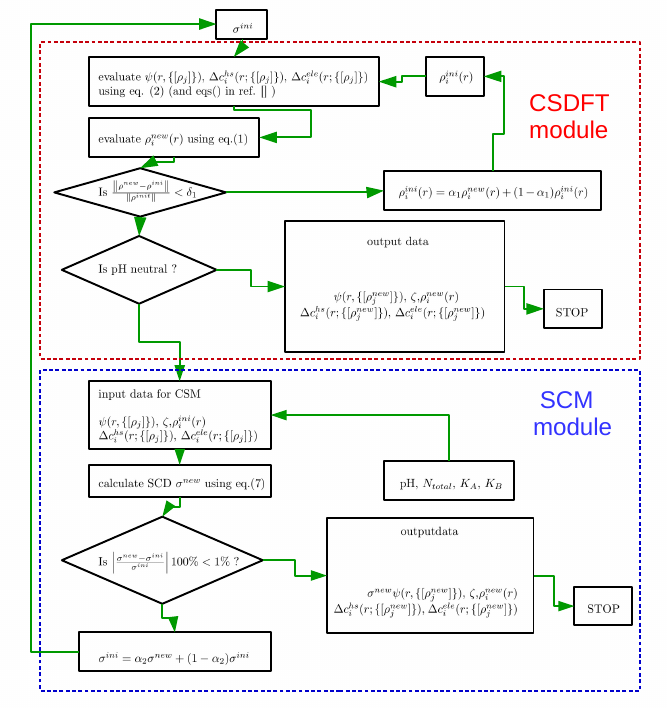}}
\par\end{centering}
\caption{Iterative Solver Scheme}
\end{figure}

\selectlanguage{english}%
Note that expression (\ref{eq:complexation}) describes the effects
of the structural and electrostatic properties of the electrolyte
on the SCD, whereas the boundary condition in expression (\ref{eq:poissoneq})
accounts for the SCD effects on the structural and electrostatic properties
of the electrolyte. Therefore, eqns (\ref{eq:ionprofilefinal})-(\ref{eq:complexation})
must be solved self-consistently as it is shown in Figure \ref{fig: solverscheme}. 

\newpage{}
\selectlanguage{american}%


\begin{thebibliography}{10}
\bibitem{key-1}Gerald S. Manning. The molecular theory of polyelectrolyte
solutions with applications to the electrostatic properties of polynucleotides.
Quarterly Reviews of Biophysics, 11(2):179\textendash{} 246, 1978.

\bibitem{key-9}David C. Grahame. The electrical double layer and
the theory of electrocapillarity. Chemical Reviews, 41(3):441\textendash 501,
1947. PMID: 18895519.

\bibitem{key-2}John Newman. Electrochemical Systems, chapter 1. Englewood
Cliffs, N.J. Prentice-Hall, 1973.

\bibitem{key-3}M. Barisik, S. Atalay, A. Beskok and S.J. Qian, J.
Phys. Chem. C 2014, 118, 1836.

\bibitem{key-4}Sonnefeld, M. Löbbus and W. Vogelsberger, Colloids
Surf. A, 2001, 195 215.

\bibitem{key-5}J.H. Masliyah and S. Bhattacharjee, Electrokinetic
and Colloid Transport Phenomena, John Wiley and Sons, Hoboken, 2006.

\bibitem{key-6}A. Studart, E. Amstad and L. Gauckler, Langmuir 2007,
23, 1081.

\bibitem{key-7}Paul A. Janmey, David R. Slochower, Yu-Hsiu Wang,
Qi Wen, and Andrejs Cebers. Polyelectrolyte properties of filamentous
biopolymers and their consequences in biological flu- ids. Soft Matter,
10:1439\textendash 1449, 2014.

\bibitem{key-8}Alexei A. Kornyshev. Double-layer in ionic liquids:
Paradigm change? Physical Chemistry B, 111(20):5545 \textendash{}
5557, 2007.

\bibitem{key-10}Ren, P., Chun, J., Thomas, D.G., Schnieders, M.J.,
Marucho, M., Zhang, J. and Baker, N.A., Quarterly Reviews of Biophysics,
2012), 49, 427.

\bibitem{key-11}N. A. Baker, D. Bashford and D. A. Case, New Algorithms
for Macromolecular Simu- lation, 2006, 49, 263.

\bibitem{key-12}Ovanesyan, Z.; Medasani, B.; Fenley, M. O.; Guerrero-García,
G. I.; Olvera de la Cruz, M. and Marucho, M. (2014). Excluded volume
and ion-ion correlation effects on the ionic atmosphere around B-DNA:
Theory, simulations, and experiments, J Chem Phys 141 : 225103 (PMCID:
PMC4265039).

\bibitem{key-13}Christian Hunley, Diego Uribe, and Marcelo Marucho,
A Multi-scale approach to describe electrical impulses propagating
along Actin filaments in both intracellular and in-vitro conditions,
submitted to RCS Advances.

\bibitem{key-14}Medasani, B.; Ovanesyan, Z.; Thomas, D. G.; Sushko,
M. L. and Marucho, M. (2014). Ionic asymmetry and solvent excluded
volume effects on spherical electric double layers: A density functional
approach, J Chem Phys 140 : 204510 (PMCID: PMC4039739).

\bibitem{key-15}Ovanesyan, Z.; Aljzmi, A.; Almusaynid, M.; Khan,
A.; Valderrama, E.; Nash, K. L. and Marucho, M. (2016). Ion-ion correlation,
solvent excluded volume and pH effects on physicochemical properties
of spherical oxide nanoparticles, J Colloid Interface Sci 462 : 325-333

\bibitem{key-16}Hunley, C. and Marucho, M. (2017). Electrical double
layer properties of spherical oxide nanoparticles, Phys. Chem. Chem.
Phys. 19 : 5396-5404

\bibitem{key-17}Jmol: an open-source Java viewer for chemical structures
in 3D. http://www.jmol.org/ 

\bibitem{key-26}Chen, C.R., and Makhatadze, M. ProteinVolume: calculating
molecular van der Waals and void volumes in proteins, , BMC Bioinformatics
2015 16:101. 

\bibitem{key-18}Voss, N.R, and Gerstein, M., 3V: cavity, channel
and cleft volume calculator and extractor, Nucleic Acids Res. 2010
July 1; 38 W555\textendash W562 

\bibitem{key-19}Dolinsky TJ, Nielsen JE, McCammon JA, Baker NA. PDB2PQR:
an automated pipeline for the setup, execution, and analysis of Poisson-Boltzmann
electrostatics calculations. Nucleic Acids Research 32 W665-W667 (2004).

\bibitem{key-20}http://www.poissonboltzmann.org/

\bibitem{key-21}Vergara-Perez, S. and Marucho, M. (2016). MPBEC,
a Matlab Program for Biomolecular Electrostatic Calculations, Comput.
Phys. Commun. 198 : 179-194

\bibitem{key-22}R. D. Shannon (1976). \textquotedbl{}Revised effective
ionic radii and systematic studies of interatomic distances in halides
and chalcogenides\textquotedbl{}. Acta Crystallogr A. 32: 751\textendash 767.

\bibitem{key-23}Warshavsky, V. and Marucho, M. (2016). Polar-solvation
classical density-functional theory for electrolyte aqueous solutions
near a wall, Phys Rev E 93 : 042607
\end{thebibliography}
\end{document}